# Music Genre Classification: Ensemble Learning with Subcomponents-level Attention

Yichen Liu, Abhijit Dasgupta, Qiwei He

*Abstract*— Music Genre Classification is one of the most popular topics in the fields of Music Information Retrieval (MIR) and digital signal processing. Deep Learning has emerged as the top performer for classifying music genres among various methods. The letter introduces a novel approach by combining ensemble learning with attention to sub-components, aiming to enhance the accuracy of identifying music genres. The core innovation of our work is the proposal to classify the subcomponents of the music pieces separately, allowing our model to capture distinct characteristics from those sub-components. By applying ensemble learning techniques to these individual classifications, we make the final classification decision on the genre of the music. The proposed method has superior advantages in terms of accuracy compared to the other state-of-the-art techniques trained and tested on the GTZAN dataset.

## I. Introduction

Audio is one of the most frequently met types of information. In the field of Music Information Retrieval (MIR), we are eager to get insights embedded inside the Music. Music created by different artists and different moods or created under different cultural backgrounds could all contain significantly different information. The tasks of MIR could include artist and music title detection, genre identification, and music transcription into symbolic representations or lyrics[1].

Music Genre Classification/Identification is one of the most vital tasks in the realm of MIR. Discovering genre information could help people better understand music, better categorize it and recommend it.

Numerous studies have been conducted by various researchers, and different techniques have been used for the development of music genre classifiers. The successful techniques include Support Vector Machines (SVM), AdaBoost, Non-negative Tensor Factorization, Sparse Representation Classification, Convolutional Neural Networks (CNN), etc.[2]

### A. Music Features for Classification

To perform meaningful classification, it is essential to extract the correct features from the music audio signal. Several features in music audio can be used for classification. Reviews conducted by Sharma et al. [3]as well as Ndou et al. [2]mentioned that timbre, rhythm, tempo, harmony, and melody are commonly applied features in traditional machine learning models. In contrast, deep learning models are capable of capturing more complex and comprehensive representations, such as spectrograms, which record both frequency and time features.

### B. Music Acoustic classification

The domain of acoustic classification has received extensive attention within the research community, with deep learning methods, such as CNN, gaining substantial prominence in recent years. While CNNs are traditionally employed for image processing, the utilization of spectrograms to represent audio signals as pixel data, subsequently processed by CNNs, has become a standard practice in Music Information Retrieval (MIR)[4]. A notable study conducted by Pelchat N et al. [5] employed a CNN on an unknown dataset comprising 15,000 150-pixel spectrograms for each genre, resulting in an 85% test accuracy. Another research done by Zhang X[6] used an RNN-LSTM network, which used the Mel-frequency cepstral coefficient as a feature that describes the brightness. The model reached an accuracy of 68.93%. Cheng et al. [7] used a pre-trained model YOLOv4 as the neural network architecture and got an accuracy of 94.5% after several iterations [9].

### C. Music Vocal Classification

Given the significant role of vocals in music compositions, an alternative approach to classification involves the discrimination of vocal elements. Several researchers have explored the utilization of vocal features to classify singers or specific aspects of vocal performances. For instance, Sha et al. [8] applied SVM to a feature set called correlation-based feature subset selection to classify singing timbre, achieving a 79.84% accuracy. Similarly, Van, T. P. et al. [9]employed Long Short-Term Memory (LSTM) networks to Mel Frequency Cepstral Coefficient feature to classify singers. However, it seems there is a lack of exploration in genre classification using vocal features.

### D. Musical Components Separation

The literature on the subject of accompaniment and vocal separation is extensive, featuring classic techniques such as nearest neighbors and median filtering methods pioneered by Fitzgerald[10]. Concurrently, modern approaches are employed, including using Deep U-Net convolutional networks, as demonstrated by Jansson[11].

### E. Ensemble Learning

Ensemble Learning was first introduced by Dieterich T. [10] in 2002. The idea of Ensemble Learning is to use multiple learners or lower-level machine learning models and combine their predictions. Ensemble Learning is widely used in classification tasks. [11]. Shariat R et al. [1] employed Ensemble Learning on the task music genre classification. In the study, the performances of applying Ensemble Learning

Yichen Liu, Abhijit Dasgupta, Qiwei He are with Georgetown University, Washington, DC 20057 USA. (e-mail:yl1520@georgetown.edu, ad1704@georgetown.edu, qh86@georgetown.edu).

varies on different dataset, since the researchers did not provide different features or sub datasets to the lower-level models.

*F. Current study*

Most researchers traditionally treat music pieces as holistic entities for classification. In contrast, Oramas et al. [4]introduced a Multimodal deep-learning model that leverages both audio data and album cover images for music genre classification[8]. This research, however, seeks to depart from convention by decomposing music compositions into two distinct components: accompaniment and vocals. A novel approach remains unexplored. The study conducted by Van et al. [9] indicates that separated vocals could perform better than raw audio signals in terms of singing voice classification, which shows that isolated vocals show some features that might not emphasized by the other components of the raw audio signal. The two classification processes of accompaniment and vocal will be executed in parallel, with the final classification decision determined through the process of Ensemble Learning.

## II. METHOD

This study used the GTZAN data set to train the model[10]. The data set contains 1000 music segments from 10 genres: blues, classical, country, disco, hip-hop, jazz, metal, pop, reggae, and rock. Each segment is a 30-second music piece with a sampling rate (SR) of 22050 Hz. In the majority of genres, except classical, disco, and jazz, music consists of two major parts: accompaniment and vocal. The vocal part refers to the singers' singing voice, while the accompaniment part refers to the instrumental sound of the music piece. The accompaniment component will be fed into a Convolution Neural Network (CNN). The vocal component will be fed into a Long Short-Term Memory (LSTM). Finally, the Ensemble Learning method will be used for the final classification decision. As shown in Fig. 1.

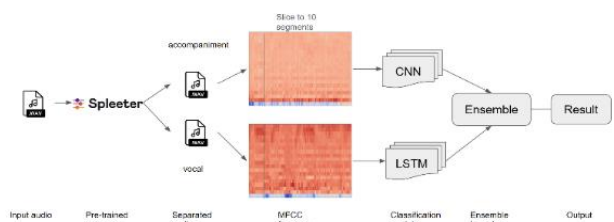

*Fig. 2 General Work Flow of Proposed Method*

*A. Data Preparation*

*1) Audio Separation*

The audios in the GTZAN dataset are whole music pieces consisting of both accompaniment and vocal, just like the music people listen to daily. The music segments will be separated into accompaniment and vocal before feeding into the neural networks for classification. Since separation is not the main task in this study, a Python package named Spleeter[11] utilizes deep U-net convolutional networks[12] is used on this task. This pre-trained model allows three different separation tasks: vocals/accompaniment separation, four stems separation (vocals, bass, drums, and other), and 5 stems separation (vocals, bass, drums, piano, and other). In this study, the separation of 2 components, vocal and accompaniment, will be performed. Moreover, due to the limited number of samples, each 30-second segment will be sliced into ten 3-second segments, which makes the dataset 10000 music pieces with labels. With the 10000 3-second music segments, we have 10000 vocals and 10000 accompaniments with labels.

*2) Feature Selection*

Each music segment will be converted into spectrograms to put the audio signals into Deep Neural Networks. Then, the Mel-frequency cepstral coefficients (MFCC) will be extracted. The MFCC feature consists of specific numbers of Mel-frequency and covers the range of frequencies of 20 Hz to 20050 Hz, which is about the human detectable frequency range[13]. Each 3-second segment will use a sliding window with a hop length of 512 milliseconds and 2048 fast Fourier transfer window size. Therefore, there are 132 frames generated per segment. 40 MFCC features will be selected as a result of a (40, 132) matrix for each segment.

*B. Lower Level Models*

*1) Vocal Classification Network*

The vocals of the music are just the singing of lyrics, which is language. Therefore, there are strong dependencies between word and word. An LSTM is a type of RNN that is designed for sequential data processing. Its ability to capture long-range dependencies in data has made LSTMs a key component in a wide range of sequential data modeling applications[14]. The vocal classifier will input the MFCC feature of each 3-second music segment's vocal component. The proposed model is a network consisting of 2 bidirectional LSTM layers with a size of 256, three fully connected dense layers with sizes of 256, 128, and 32, and one output layer, shown in Fig. 2. The fully connected layers used Relu as the activation function, and the output layer used softmax as its activation function. The output layer will provide a 10-variable vector as the output. Each variable shows the probability of each label.

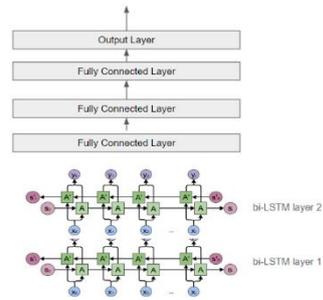

*Fig. 1 Architecture of LSTM Model. Adapted from* [15].

*2) Accompaniment Classification Network*

The accompaniment of music is much more complicated than the vocal. The accompaniment could consist of different types of instruments, and the sequence of a particular kind of instrument could be periodic or aperiodic. CNN is a neural network originally designed to analyze visual data, such as images and videos. It employs convolutional layers to automatically learn and detect features like edges, textures, and patterns. Therefore, a CNN model is chosen to capture the general characteristics of the MFCC spectrogram. The accompaniment classifier inputs the MFCC feature of each 3-second music segment's accompaniment component. The model used is a CNN Fig. 3. The network consists of 4 two-dimensional convolution layers with a two-dimensional max pooling layer after each convolution layer. The convolution layers have sizes of 64, 32, 32, 16. There is a fully connected dense layer after flattening the 2-dimensional data into 1-dimensional. All of the convolution layers and fully connected layers use Relu as their activation function. There is another fully connected layer using the softmax activation function as the output layer, which generates a 10-variable vector showing the probability of each label.

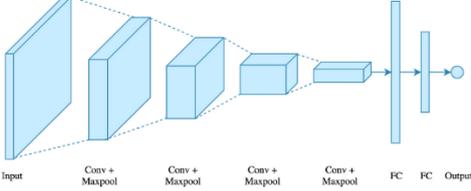

*Fig. 3 Architecture of CNN Model. Adapted from* [16].

## C. Ensemble Learning

With the prediction from vocal classification and accompaniment classification networks, ensemble learning techniques will be used to improve the performance by combining the strength of the two base models. Two standard ensemble learning techniques are chosen: Bagging and Stacking [17].

### 1) Bagging

Bagging is a method that combines the predictions of ensemble numbers using simple statistics, such as voting or averaging. Soft voting combines the results of all classifications with a certain weight. Due to the different characteristics of different genres, different weights might be used for different genres. For instance, the genre "classical" usually does not have any vocals, and the classification results from vocals for classical music might not be considered, which means the weight for vocals in classical music will be set to zero. Averaging is a method that finds the average value of each prediction.

The prediction from the vocals classification is denoted as

$$x_v = [x_v^0, x_v^0, \ldots, x_v^{i-1}, x_v^i]$$

where $i$ is the label of genres, $i \in [0,9]$

The prediction from the accompaniment classification is denoted as

$$x_a = [x_a^0, x_a^0, \ldots, x_a^{i-1}, x_a^i]$$

where $i$ is the label of genres, $i \in [0,9]$

The final classification decision denoted as

$$y = [y^0, y^0, \ldots, y^{i-1}, y^i]$$

where $i$ is the label of genres, $i \in [0,9]$

The proposed combining method includes ignoring vocal/accompaniment predication:

$$y^i = x_a^i * w_a + x_v^i * w_v$$
where $w_v = 0$ or $w_a = 0$

Soft voting:

$$y^i = x_a^i * w_a + x_v^i * w_v$$

Mean value

$$y^i = x_a^i * w_a + x_v^i * w_v$$

where $w_a, w_v = 0.5$

With the final classification decision array $y$, $y^i$ with the highest value indicates the classification label.

### 2) Stacking method

The stacking method introduced the idea of the meta-model, which means a higher-level model that uses the output of the base models (or lower-level model) as the input and learns from the input. The output of the meta-model is the final prediction decision[17]. As Sewell noted.'The procedure is as follows:
1. Split the training set into two disjoint sets.
2. Train several base learners on the first part.
3. Test the base learners on the second part.
4. Using the predictions from 3) as the inputs and the correct responses as the outputs, train a higher-level learner.[17]'

The proposed stacking method includes Linear Regression, XGBoost, and Neural Network. In this study, the input of each meta-model will be an array with a length of 20, which consists of two arrays with a length of 10 from the output of vocal and accompaniment predictions.

## III. RESULT

### A. Hyperparameter Tunings on LSTM and CNN Models

A series of hyperparameter tunings was performed to increase the performance of the models. Dropout layers, L1 and L2 regularizations were used to overcome outfitting. Different learning rates were tested to reach the best performance. Early stopping techniques were also used to prevent the overfitting situation. Most importantly, different combinations and numbers of convolution, LSTM, and Dense layers were tested. The final performance of the classification model of accompaniment segmentation using CNN reached an accuracy of 74.6%. The final performance of the classification model of vocals segmentation using LSTM reached an accuracy of 61.4%.

### B. Evaluations of LSTM and CNN Models

Fig. 4 and Fig. 5 are the confusion matrixes of accompaniments and vocals prediction. Labels 0 – 9 indicate ten genres (blues, classical, country, disco, hip-hop, jazz, metal, pop, reggae, and rock).

The classifications of these two subcomponents indicate that the model is sensitive to overlapping characteristics within these genres. However, the two classification results show their strengths in classifying different genres, which makes it necessary and meaningful to ensemble the results.

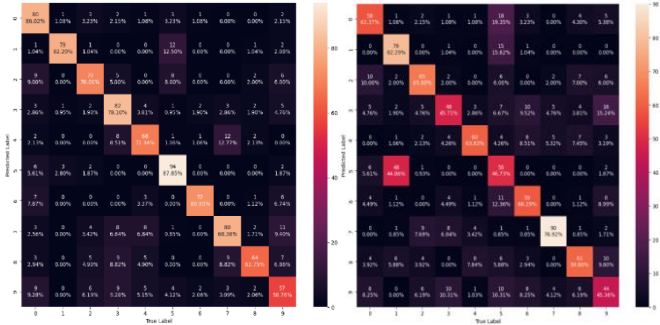

Fig. 4 Confusion Matrix of Classification of Accompaniments

Fig. 5 Confusion Matrix of Classification of Vocals

## C. Evaluation of Ensemble Learning

A total of five different methods of fusing were used: XGBoost, Logistic Regression, Dense Neural Network, Mean Averaging, and Soft Voting with the evaluation based on Recall, Precision, Accuracy, and F1 Score, shown in the Table I. The performance of all ensemble learning methods could all outperform the individual models. Stacking Ensemble methods performed better than others. Logistic Regression showed the most balanced performance across all metrics, slightly surpassing the XGBoost and ANN approaches by 0.3% and 0.4% F1 score and accuracy, suggesting a solid alignment of this model with the task's requirements. Notably, the Bagging Ensemble methods, Mean Averaging, and Soft Voting did not perform as well as the Stacking Ensemble techniques since the weight for voting of each class was challenging to optimize. The best-performing bagging ensemble method is the mean averaging method, which outperformed the soft voting method by 0.6% accuracy and 0.5% F1 score.

When we used the same input feature spectrogram or MFCC, by comparing with the models proposed by other researchers in Table II [19] – [21], the proposed Ensemble Learning method could outperform all. The study uses a YOLOv4 pre-trained model that performed extremely well, with an accuracy of 94.5%. Some researchers used further selected features as input for the models. For example, Fu Z et al. [21] used a combination of 8 types of individual features, includes timbre features based and temporal features based MFCC, Amplitude Spectrum Envelop (ASE), and Octave based Spectral Contrast (OSC) as well as beat and chord, as the input in their study. The input increased the general complexity of the model, but it did increase the performance of the model. The SVM model proposed provided an accuracy of 90.95%. Meanwhile, the same SVM model that uses a single MFCC as the input offers an accuracy of 78.92%. Another study used an SVM model with 606 further selected features that reached an accuracy of 81.9%.

TABLE II
Accuracy for Different Ensemble Models

| Feature Type | Model | Accuracy |
|---|---|---|
| Lower-Level Musical Feature (Spectrogram) | CNN[19] | 73.3% |
|  | CNN[20] | 74% |
|  | YOLOv4[7] | 94.5% |
|  | SVM[21] | 78.92 |
|  | **CNN+LSTM+LR** | **82%** |
| Higher-Level Musical Feature (Further selected features) | SVM[18] | 90.85 |
|  | SVM[22] | 81.9 |

## IV. CONCLUSION AND DISCUSSION

In short, the study proves that you can classify music genres more accurately by paying attention to subcomponents. Compared to methods proposed by other researchers, the proposed method could outperform the state-of-the-art deep learning models using lower-level audio features. A research used pre-trained YOLOv4 as the classification model also provides a new idea of building the model[7]. It's not very meaningful to compare our shallow self-trained model with the complex pre-trained model, but we could see the effectiveness of this idea. Therefore, employing a pre-trained model could also be one of the future works. Another future study is that more subcomponents, such as drums, bass, and piano, could be extracted.

In general, the proposed method could provide unique insights from the subcomponents level, which could be helpful for things like recommending music and organizing digital music collections. At the same time, the idea of subcomponents-level attention could be applied to other forms of media or even combined with the concept of Multimodal Learning. For example, a video recommendation system with attention on characters or objects exists in the video or the background music.

## REFERENCES

[1] R. Shariat and J. Zhang, "An Empirical Study on the Effectiveness of Feature Selection and Ensemble Learning Techniques for Music Genre Classification," in *Proceedings of the 18th International Audio Mostly Conference*, New York, NY, USA, 2023, pp. 51–58.
[2] N. Ndou, R. Ajoodha, and A. Jadhav, "Music Genre Classification: A Review of Deep-Learning and Traditional Machine-Learning

TABLE I
Accuracy, Recall, Precision, F1 Score for Different Ensemble Models

| Type | Model | `Recall | Precision | Accuracy | F1 Score |
|---|---|---|---|---|---|
| **Stacking Ensembles** | XGBoost | 81.7% | 81.9% | 81.7% | 81.8% |
|  | **Logistic Regression (LR)** | **82%** | **82.4%** | **82%** | **82.1%** |
|  | Dense Neural Network | 81.6% | 82.2% | 81.6% | 81.7% |
| Bagging Ensembles | Mean Averaging | 80.2% | 80.2% | 80.1% | 80% |
|  | Soft Voting | 79.5% | 80% | 79.5% | 79.5% |
| Individual Models | CNN on Accompaniments | 74.7% | 75.9% | 74.6% | 74.8% |
|  | LSTM on Vocals | 61.4% | 62.5% | 61.4% | 61.5% |